\documentclass[twocolumn,showpacs,preprintnumbers]{revtex4}%   
\usepackage{amsmath}   
\usepackage[truetex]{graphicx}   
\usepackage{subfigure}   
\usepackage{bm}   
\usepackage{amsfonts}   
\usepackage{amssymb}%   
\begin{document}   
\title{Propagation and Backscattering of Mechanical Impulses in a Gravitationally
Loaded Chain: Dynamical Studies and Toy Model Based Phenomenology
}
\author{Edgar \'{A}valos}
\email{edgar.avalos@excite.com}   
\author{T. R. Krishna Mohan}
\email{kmohan@cmmacs.ernet.in}   
\altaffiliation{On leave from CSIR Centre for Mathematical Modelling and Computer Simulation, Bangalore 560037, India}
\author{Surajit Sen}
\thanks{Corresponding author}
\email{sen@dynamics.physics.buffalo.edu}   
\affiliation{Department of Physics, State University of New York. Buffalo, New York   
14260-1500}   
\date{\today}   
   
\begin{abstract}   
We recently introduced a simple toy model to describe   
energy propagation and backscattering in complex layered media   
(T.R. Krishna Mohan and S. Sen, Phys. Rev. E {\bf 67}, 060301(R) (2003)).   
The model provides good qualitative description of energy propagation and    
backscattering in real soils. Here we present a dynamical study of energy propagation   
and backscattering in a gravitationally loaded granular chain and   
compare our results with those obtained using the toy model.  
The propagation is ballistic for low $g$ values and acquires   
characteristics of acoustic propagation as $g$ is increased. We focus on the   
dynamics of the surface grain and examine the backscattered energy at the surface.  
As we shall see, excellent agreement between the two models is achieved when we consider   
the simultaneous presence of acoustic and nonlinear behavior in the toy model.  
Our study serves as a first step towards using the toy model to describe   
impulse propagation in gravitationally loaded soils.    
\end{abstract}   
\pacs{46.40.Cd,45.70.-n,43.25.+y}
\maketitle

\section{\label{sec1}Introduction}   
   
Acoustic imaging of buried objects in a non-linear medium like nominally dry soil is still   
an open problem \cite{senburied}. Amongst its main applications, we can   
mention the problem of locating antipersonnel land mines, human remains for   
forensic investigations, hidden underground structures of archaeological   
importance etc. It has been shown that gentle mechanical   
impulses~\cite{rogdon} can be used to detect buried objects at depths of a   
meter or so in nominally dry sand beds. Imaging requires the understanding of   
pulse propagation in nonlinear media. Besides the underlying non-linearity due   
to the Hertzian contact forces between the grains (see,   
for example, \cite{Landau 1970}), gravitational loading is an added feature   
that affects the dynamics. Before we can analyze the gravitational effects in   
3D systems, it is reasonable first to look at the problem in 1D. Further   
motivation for 1D studies can be found in \cite{mosen}.   
   
\begin{figure}[ptb]   
\centering   
\includegraphics[scale=0.43]{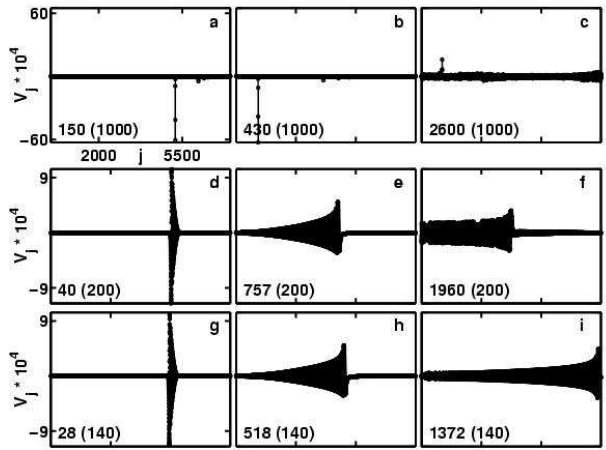}  
\caption{In Fig.~\ref{prop}%   
(\emph{a})-(\emph{c}), $g=0$ and we see solitary wave propagation. In   
Fig.~\ref{prop}(\emph{d})-(\emph{f}), $g=0.1$ and in Fig.~\ref{prop}%   
(\emph{g})-(\emph{i}), $g=1.0$. In both cases, open boundary conditions are   
employed at the top with closed boundary conditions at the bottom. We see   
solitary wave like front with a long tail for non-zero $g$. The text in the   
lower left hand corner indicates the times at which the snapshots are taken   
and, in brackets, is given the roundtrip time, i.e. the time taken by the   
pulse to travel up and down the chain once. The $y$-axis ranges have been   
marked in the leftmost panel for each row of panels while the $x$-axis range   
is the same in all and labelled in the top left panel, Fig.~\ref{prop}%   
(\emph{a}), where $j$ is the position of the particle in the chain and $V_{j}$   
its velocity.}%   
\label{prop}%   
\end{figure}%  
Impulse propagation in a gravitationally loaded chain with $N$   
grains and a perfectly reflecting boundary at the bottom has been studied \cite{Sinkovits 1995, Sen 1996, nest}.   
When $g=0$, one finds solitary wave propagation \cite{Ne83, Sinkovits 1995} as shown via velocity vs.   
position plots made at different times in Figs.~\ref{prop}(\emph{a}%   
)-(\emph{c})\ (see later for simulation details). In all the panels of   
Fig.~\ref{prop}, the numbers in the bottom left corner indicate the times at   
which the snapshots have been taken and, in brackets, the round trip time of   
the pulse is given.   
   
The velocity of the solitary wave is a function of   
its amplitude \cite{Ne83, Sinkovits 1995, Manciu 1999 III, SenManciu pre 2001}. Each solitary   
wave is reflected at the bottom boundary. During reflection, the solitary wave   
forms secondary solitary waves \cite{Manciu pre 2001, JobSen,   
quasi}. Further, secondary solitary waves are generated from collisions with other solitary waves \cite{mosen}. The formation of secondary solitary waves progressively reduces the amplitude of   
the solitary waves in the system.  In   
Fig.~\ref{prop}(\emph{c}), we find that the amplitude of the primary solitary   
wave has diminished considerably after $2.6$ roundtrips.   
  
When $g>0$, one finds a solitary wave like front with a tail that elongates   
with time \cite{Hong 2001, Hong 1999, Manciu 2000}. In Figs.~\ref{prop}(\emph{d})-(\emph{f}), we have shown snapshots from the   
propagation of the impulse in a system with $g=0.1$ and with open boundary   
condition at the top. We have the same boundary conditions in Figs.~\ref{prop}%   
(\emph{g})-(\emph{i})\ with $g=1.0$. We see that the impulse propagates faster with increased $g$.   
  
The scheme of this paper is as follows. Section II surveys briefly the $1$%   
-$D$\ model with the simulation details. Our focus is to probe impulse   
propagation and backscattering by simply considering the dynamics of the   
surface grain. In Section III, we construct a version of the toy model and   
recover the results obtained in section II. Section IV summarizes the conclusions for the $1$D study and assesses the usefulness of this study for further $3$D analysis.   
   
\section{Simulation model}   
   
We model the granular chain as a collection of $10000$ spherical grains   
which are placed in contact with one another and loaded by a gravitational   
field. The interaction between every pair of spheres of radius $R$ is   
driven by Hertz law \cite{Landau 1970, Sinkovits 1995},\textit{\ i.e.}%   
, it is assumed that spheres labeled as $i$ and $i+1$ are   
interacting with a potential proportional to the overlaps, $\delta_{i,i+1}$,   
in the contact region,   
\begin{equation}   
V(\delta_{i,i+1})=   
\begin{cases}   
a\ \delta_{i,i+1}^{5/2} & \text{if $r_{i,i+1} \leq 2R$}\\   
0 & \text{if $r_{i,i+1} > 2R $}%   
\end{cases}   
\label{Hertz potential}%   
\end{equation}   
where $a$ is a constant that depends upon the material characteristics of the   
grains, $r_{i,i+1}$ is the separation distance between the centers of the grains $i   
$ and $i+1$ and $\delta_{i,i+1}\equiv   
2R-r_{i,i+1}$ is the grain overlap. The equation of motion for grain $i$   
can be written as \cite{Manciu 2000},%   
\begin{equation}   
m\overset{\cdot\cdot}{z_{i}}=\frac{5}{2}a\left(  \delta_{i,i-1}^{3/2}%   
-\delta_{i+1,i}^{3/2}\right)  -mg\text{,}%   
\end{equation}   
where $m$ and $g$ are the mass of the grain and the value of gravity,   
respectively. The system dynamics is obtained by time integration of the   
coupled Newtonian equations of motion via the velocity-Verlet algorithm   
\cite{Tildesley 1987}. We ignore the role of dissipation in the present study; dissipation effects can be built in later \cite{diss}.   
We set $2R$ and $m$ to $1$, and $a$ is set equal to   
$5657$. We find that an integration time step of $1\times 10^{-5}$ provides excellent enrgy conservation; decreasing the time step further does not improve the accuracy of our results. Particles are labeled starting from the bottom so that the   
$N$$^{\mbox   
{\small th}}$\ particle is at the top. While the bottom boundary condition has   
been kept perfectly reflecting in all the simulations, both open and closed   
boundary conditions have been employed at the top in different simulations to   
investigate the differences induced by the particular choice.   
   
\begin{figure}[ptb]   
\centering   
\includegraphics[scale=0.45]{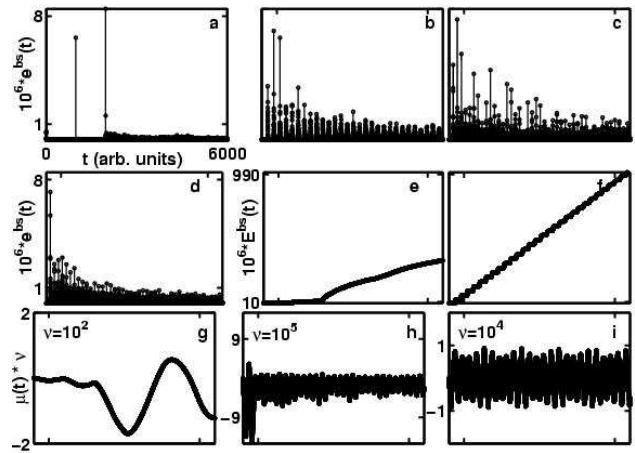}  
\caption{In Fig.~\ref{bcksc}%   
(\emph{a}), $g=0$, in Fig.~\ref{bcksc}(\emph{b})\ $g=0.1$, and in   
Figs.~\ref{bcksc}(\emph{c})-(\emph{d}), $g=1.0$. Figs.~\ref{bcksc}%   
(\emph{a})-(\emph{c})\ have closed boundary conditions at the top while   
Fig.~\ref{bcksc}(\emph{d})\ has open boundary condition at the top. Typical   
integrated backscattered energy, $E^{bs}(t)$, plots are shown in   
Figs.~\ref{bcksc}(\emph{e})-(\emph{f})\ where Fig.~\ref{bcksc}(\emph{e})\ is   
for $g=0.0$ and Fig.~\ref{bcksc}(\emph{f})\ for $g=0.1$ with top closed   
boundary condition (cf. Fig.~\ref{bcksc}(\emph{b})). The center of mass, $\mu   
$, oscillations typically found in these systems are given in the third row of   
panels. Fig.~\ref{bcksc}(\emph{g})\ is for $g=0.0$ while Fig.~\ref{bcksc}%   
(\emph{h})\ is for the top closed boundary condition ($g=0.1$; cf.   
Fig.~\ref{bcksc}(\emph{b})\ and Fig.~\ref{bcksc}(\emph{f})). 
Fig.~\ref{bcksc}(\emph{i})\ is for the top open boundary condition (cf.   
Fig.~\ref{bcksc}(\emph{d})). The $x$-axis remains the same in all panels as   
labelled in Fig.~\ref{bcksc}(\emph{a})\ with the time, $t$, given in arbitrary   
units (a.u.) of the simulation. $y$-axis ranges for the first row is given in   
the leftmost panel, and separately labelled for each panel in the bottom row.   
For the middle row, the first panel on the left has a different $y$-axis range   
as shown and, for the last two panels, it is as shown for the middle panel.}%   
\label{bcksc}%   
\end{figure}  
The dynamics is initiated via an initial velocity perturbation at   
the top of the chain, $v_{N}(t=0) = v_{in}$, with $v_{i}(t=0) = 0$ for $0 \le i <   
N$; we have employed $v_{in}=-0.01$ in the studies reported here. We   
subsequently monitor the backscattered energy received at the surface,   
$e^{bs}(t) = v_{N}^{2}(t)/2$. The time integrated   
backscattered energy, $E^{bs}(t)$, has also been computed by summing the   
entire sequence of energy packets received at the surface, i.e.   
$E^{bs}(t) = \sum_{t^{\prime}=0}^{t} e^{bs}(t^{\prime})$.   
   
In Fig.~\ref{bcksc}(\emph{a}-\emph{d}), we show plots of $e^{bs}(t)$ vs. $t$.   
Fig.~\ref{bcksc}(\emph{a})\ is obtained with $g=0.0$. We have two prominent   
peaks at early times followed by a continuous distribution of much smaller   
peaks. Clearly, the smaller peaks originate in the breakdown of the solitary waves at the boundaries and through mutual interactions (this issue is discussed in detail in \cite{quasi}). The two large peaks at early times are separated by roundtrip times of the system.  
Fig.~\ref{bcksc}(\emph{b})\ corresponds to $g=0.1$,   
with reflecting boundary conditions at the top in which the surface grain is only allowed to move in one direction into the chain. In this case, there are large   
peaks separated by roundtrip times of the system, but there is much activity in between them   
indicating the presence of acoustic like rattling throughout the system.   
Increasing $g$ to $1.0$ (Fig.~\ref{bcksc}(\emph{c})), with the same boundary   
conditions, does not reveal any noticeable change in the pattern except for   
increasing the density of peaks, caused by the smaller roundtrip times. If we keep $g = 1.0$ and change to   
open boundary conditions, meaning that the surface grain is allowed to move up and down, we get the pattern shown in   
Fig.~\ref{bcksc}(\emph{d})\ where we notice that the number of larger peaks   
are reduced. We attribute this to increased energy sharing caused by the   
open boundary at the top.   
   
The $E^{bs}(t)$ vs. $t$ plots show a distinct difference between   
the ballistic and acoustic type propagation cases. While Fig.~\ref{bcksc}%   
(\emph{e})\ is for the $g=0$ case, Fig.~\ref{bcksc}(\emph{f})\ is typical of   
$g>0$ cases (here,   
$g=0.1$ with closed boundary   
at the top). The increase is linear for non-zero $g$ whereas it only   
approaches linear for $g=0.0$ in the later stages. The steps in $E^{bs}%   
(t)$\ indicate the arrival of peaks with the quiescent period in between   
marked by plateaus; the length of the plateau indicates the time period   
between peaks.   
   
The pattern of arrival of large peaks introduce modulations in the amplitudes   
of $e^{bs}(t)$. Nevertheless, there are no simple patterns   
in the distribution of peaks of $e^{bs}(t)$ vs. $t$ in Figs.~\ref{bcksc}(\emph{a}-\emph{d}).   
We note that the boundary conditions affect $e^{bs}(t)$.   
A possible modulating mechanism could be due to the center of mass  
(com) oscillations of the system.  
We show these   
oscillations (the \textquotedblleft breathing\textquotedblright\ of   
the chain) in the third row of Fig.~\ref{bcksc} for three typical cases.   
With $g=0$ (Fig.~\ref{bcksc}(\emph{g})), the significant   
com oscillations appear only after the solitary waves have broken down.  
For non-zero $g$, the com oscillations are affected by the surface boundary.   
Fig.~\ref{bcksc}(\emph{h})\ is for $g=0.1$ and   
with closed boundary conditions at the top    
while Fig.~\ref{bcksc}(\emph{i})\ is   
typical of open boundary conditions at the top (in this case, $g=1.0$).  
We now turn to the toy model to see whether   
these patterns in $e^{bs}(t)$\ can be reproduced.   
   
\section{Comparison with toy model}   
   
We had introduced a toy model in an earlier paper where the energy propagation   
in a vertical alignment of masses was considered in a simplified   
manner~\cite{mosen}. At time $t=0$, we set initial energy $E=1$ for layer one   
and zero for the rest. At $t = 1$, the first layer in the vertical chain   
transfers $p$ ($<1$) of the impulse energy to the second layer, and retains   
$(1 - p)$. At subsequent times, the impulse will propagate in the same fashion   
all the way down the chain. After each transfer of energy, the phase of the   
mass reverses so that it will interact with its adjacent layer in the opposite   
direction in the following time step. The interaction between two adjacent   
layers occurs in the following two ways: (i) \textit{equipartition} case, and   
(ii) \textit{exchange} case. In the \textit{equipartition} case, the two   
interacting layers will come away from the interaction with equal amounts of   
energy; we add up the individual energies of the two layers and divide the sum   
equally between them. In the \textit{exchange} case, we let the layers   
exchange their energies; the two interacting layers, after the interaction,   
come away with the energy of the other.   
   
\begin{figure}[ptb]   
\centering   
\includegraphics[scale=0.45]{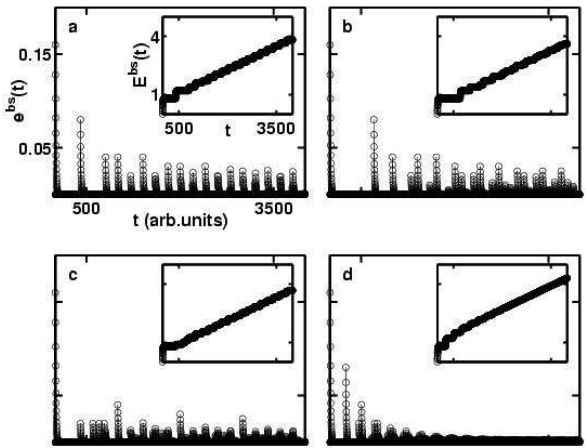}  
\caption{All the panels along with the   
insets have the axes ranges as labelled in the top left panel. Fig.~\ref{fig3}%   
(\emph{a})\ is for the case of a single equipartition center in the toy model   
located at particle no. 200 out of a total of 500 particles in the chain, and   
Fig.~\ref{fig3}(\emph{b})\ for the case, similarly, with equipartition center   
at particle no. 350; the exchange coefficient $p=0.8$. Data for these two   
cases have been averaged, along with data for cases with single equipartition   
centers located at 300 and 390 respectively, and presented in Fig.~\ref{fig3}%   
(\emph{c}). In Fig.~\ref{fig3}(\emph{d}), we have multiple equipartition   
centers located at particle nos. 125, 230 and 390 with $p=0.8$.}%   
\label{fig3}%   
\end{figure}  
In Fig.~\ref{fig3}, we show the results where the model has, for the first time, both   
equipartition and exchange. Our goal is to develop the propagation patterns seen in Figs.~\ref{bcksc}(\emph{b}-\emph{d}). The model is based on the exchange   
case, but we allow equipartition centers to develop, as time goes by, at   
different points in the chain. In Fig.~\ref{fig3}(\emph{a}), we have shown a   
case where a single equipartition center is kept fixed at layer number 200   
in a chain of 500 layers.   
Fig.~\ref{fig3}(\emph{b})\ is for a similar case but with the equipartition   
center located at layer number 350. We see that the patterns in $e^{bs}%   
(t)$\ do change slightly depending on the location of the isolated   
equipartition center, but, in both cases, it is seen that the patterns are   
broadly similar to the non-zero $g$ cases seen in Figs.~\ref{bcksc}%   
(\emph{b})-(\emph{d}).   
  
This has prompted us to try averaging over the patterns   
resulting from differing locations of the isolated equipartition centers and,   
indeed, the plot of Fig.~\ref{fig3}(\emph{c})\ shows that such averaging does   
retain the similarity with plots of Figs.~\ref{bcksc}(\emph{b})-(\emph{d}). To   
obtain this figure, we averaged over, along with the cases given in   
Fig.~\ref{fig3}(\emph{a})\ and Fig.~\ref{fig3}(\emph{b}), two other cases with   
isolated equipartition centers, in each case, located at layer numbers 300   
and 390 respectively. If we vary the exchange coefficient $p$, we get similar   
patterns (not shown here) but with the difference that the growth in   
$E^{bs}(t)$\ is proportional to the $p$ value. This indicates that $p$ could   
also be used, along with the equipartition centers, to capture the transition   
from ballistic propagation to acoustic propagation. In Fig.~\ref{fig3}%   
(\emph{d}), we show results with multiple (three, in this cases) equipartition   
centers in the system, at layer numbers 125, 230 and 390; $p=0.8$ for this   
case. We see that $E^{bs}(t)$\ grows by as much as in Figs.~\ref{fig3}%   
(\emph{a})-(\emph{c})\ but the attenuation in peaks of $e^{bs}(t)$\ is very   
pronounced and similar to the pattern seen in Fig.~\ref{bcksc}(\emph{d}).   
   
\section{Summary and conclusions}   
   
In this paper, we have studied the problem of impulse propagation in a   
gravitationally loaded granular chain. We have focused our attention   
on the backscattered energy received at the surface after an impulse has been   
initiated into the system. Our studies have been carried out using two different  
approaches--- (i) using Newtonian dynamics in a non-disspative system to describe   
backscattering at the surface of the system and, (ii) using the toy model \cite{mosen}   
to recover the behavior in (i).  
  
Our results show that the toy model is capable of reproducing the   
correct form of the backscattered energy as a function of time in   
a gravitationally loaded chain. Gravitational loading introduces   
acoustic-like oscillations in the system. Such oscillations vanish   
and the system ends up propagating solitary waves when gravitational   
loading is zero. To achieve these descriptions, one must generalize the   
earlier version of the toy model \cite{mosen}   
and incorporate both exchange (represents non-linear) and equipartition (represents acoustic)   
effects. We find that the role of the equipartition (acoustic) effect in the toy   
model is rather dominant and only limited equipartitioning of energy   
gives the correct backscattering behavior. Our studies confirm that   
mechanical propagation in a granular chain is strongly nonlinear even in   
the presence of gravitational loading.

\section{Acknowledgment}  
  
EA acknowledges the Fulbright Foundation for support. TRKM and SS   
have been supported by the Army Research Office and NASA.

\end{document}